\documentclass[superscriptaddress,altaffilletter,preprint]{revtex4-1}
\usepackage{graphicx,amsmath}
\usepackage{color}

\usepackage{bm}

\begin{document}
\title{Optical Control of Chiral Charge Pumping in a Topological Weyl Semimetal }

\author{M. Mehdi Jadidi}
\email{mehdi.jadidi@columbia.edu}
\affiliation{Department of Applied Physics and Applied Mathematics, Columbia University, New York, NY 10027 USA}

\author{Mehdi Kargarian}
\affiliation{Department of Physics, Sharif University of Technology, Tehran 14588-89694, Iran}

\author{Martin Mittendorff}
\affiliation{Institute for Research in Electronics \& Applied Physics, University of Maryland, College Park, MD 20742, USA}
\affiliation{ Universit{\"a}t Duisburg-Essen, Fakult{\"a}t f{\"u}r Physik, 47057 Duisburg, Germany}

\author{Yigit Aytac}
\affiliation{Institute for Research in Electronics \& Applied Physics, University of Maryland, College Park, MD 20742, USA}
\affiliation{Science Systems and Applications, Inc. Lanham, MD 20706, USA}

\author{Bing Shen}
\affiliation{Department of Physics and Astronomy, University of California, Los Angeles, CA 90095, USA}

\author{Jacob C. K\"onig-Otto}
\affiliation{Helmholtz-Zentrum Dresden-Rossendorf, PO Box 510119, D-01314 Dresden, Germany}

\author{Stephan Winnerl}
\affiliation{Helmholtz-Zentrum Dresden-Rossendorf, PO Box 510119, D-01314 Dresden, Germany}

\author{Ni Ni}
\affiliation{Department of Physics and Astronomy, University of California, Los Angeles, CA 90095, USA}

\author{Alexander L. Gaeta}
\affiliation{Department of Applied Physics and Applied Mathematics, Columbia University, New York, NY 10027 USA}

\author{Thomas E. Murphy}
\affiliation{Institute for Research in Electronics \& Applied Physics, University of Maryland, College Park, MD 20742, USA}

\author{H. Dennis Drew}
\email{hdrew@umd.edu}
\affiliation{Center for Nanophysics and Advanced Materials, University of Maryland, College Park, Maryland 20742, USA}

\begin{abstract}  
Solids with topologically robust electronic states exhibit unusual electronic and optical transport properties that do not exist in other materials. A particularly interesting  example is chiral charge pumping, the so-called chiral anomaly, in recently discovered topological Weyl semimetals, where simultaneous application of parallel  DC electric and magnetic fields creates an imbalance in the number of  carriers of opposite topological charge (chirality). Here, using time-resolved terahertz measurements on the Weyl semimetal TaAs in a magnetic field, we optically interrogate the chiral anomaly by dynamically pumping the chiral charges and monitoring their subsequent relaxation. Theory based on Boltzmann transport shows that the observed effects originate from an optical nonlinearity in the chiral charge pumping process. Our measurements reveal that the chiral population relaxation time is much greater than 1 ns. The observation of terahertz-controlled chiral carriers with long coherence times and topological protection suggests the application of Weyl semimetals for quantum optoelectronic technology. 

\end{abstract}%

\keywords{Dirac and Weyl semimetals, Optics, Chiral anomaly,  topological effects, pump-probe}
\maketitle

\section*{Introduction}

The control of quantum matter with light is at the forefront of  condensed matter physics research.  Recently, strong optical pumping has been employed to generate exotic electronic states  in solids,   not present at equilibrium, such as light-induced superconductivity \cite{mitranoPossibleLightinducedSuperconductivity2016a}, 
nonlinear Hall effect \cite{maObservationNonlinearHall2019a}, and ultrafast symmetry switch \cite{sieUltrafastSymmetrySwitch2019a}.  Among the recently discovered solids, Weyl semimetals  have attracted attention because they are predicted to exhibit a host of novel  topological properties not seen in other materials  \cite{burkovWeylSemimetalTopological2011c,hosurChargeTransportWeyl2012c,armitageWeylDiracSemimetals2018c}. In the electronic band structure of these materials, Weyl fermions with opposite topological charge (chirality) are separated in momentum space around the band-touching points called ``Weyl nodes''  that are sources and sinks of Berry curvature \cite{burkovWeylSemimetalTopological2011c,hosurChargeTransportWeyl2012c,armitageWeylDiracSemimetals2018c}.    An example of a topological Weyl fermions transport effect is the  chiral anomaly, where simultaneous application of parallel static electric and magnetic fields pumps the carriers from one Weyl node to the other,  unbalancing the number of chiral carriers \cite{zyuzinTopologicalResponseWeyl2012a,nielsenAdlerBellJackiwAnomalyWeyl1983c}. Since the Weyl nodes of oppsite chirality are separated in momentum space, the relaxation of chiral anomaly is expected to be slow, similar to long inter-valley scattering of carriers in 2D semiconductors \cite{rivera2016valley}. Recently,  terahertz-excited  Weyl fermions via chiral anomaly in a Weyl semimetal were proposed as an attractive platform for qubits with large coherence time to gate time ratio \cite{kharzeev2019}.  However, terahertz excitation of chiral anomaly and  the chiral population  relaxation time in a Weyl semimetal  have yet to be measured. 

Here we demonstrate optical (at terahertz frequencies) excitation and control of chiral anomaly. We employ terahertz pump-probe measurements in a magnetic field on the Weyl semimetal tantalum arsenide (TaAs) to optically modify the dynamical chiral charge pumping and monitor its subsequent relaxation. In these measurements, a strong quasi-continous wave  pulse (pump)  at 3.4 THz (14 meV ) polarized parallel to an applied magnetic field  modifies the chiral current. The change in the chiral current is monitored via reflection measurement of a second, co-polarized weak pulse (probe) at the same photon energy as pump as a function of the time delay between the two pulses. In addition to the expected fast response of hot carrier relaxation, we observe a metastable response ($\gg 1$ ns)  that we associate with the pump-induced change in the dynamical chiral current. Our study of the chiral anomaly circumvents the problems arising from electrical contacts and current jetting that have plagued electrical magneto-resistance measurements  \cite{liangExperimentalTestsChiral2018c}. By using low-energy photons in the terahertz domain, we ensure that only carriers near the Weyl pockets enclosed the Berry curvature monopoles are excited and studied-- a condition that has not been met in prior experimental studies reported to date  \cite{wuGiantAnisotropicNonlinear2017c,maDirectOpticalDetection2017a,osterhoudt2019colossal,ma2019nonlinear}.  Our measurements at different magnetic fields, pump fluences, and pump polarizations  disentangle the chiral anomaly signals from the pump-induced transient hot carriers effects. In addition to the chiral population relaxation time, these experiments yield the electron cooling rate as a function of magnetic field and polarization in TaAs, which varies widely due to phase space restrictions for scattering between Landau levels. We present a theory, based on the Boltzmann transport equations in the presence of a strong driving optical field, that explains the observed slow relaxation rate. 

In the current manuscript,  we introduce the concept of  the dynamical chiral anomaly and its optical control in a Weyl semimetal. We present the pump-probe results at zero magnetic field are presented and conclude that the response is attributed to the pump-induced hot carriers in TaAs. We show the pump-probe measurement results at various  magnetic fields applied parallel to pump$/$probe polarization, and for different pump fluences, and the results  illustrate the ability to optically control the dynamical chiral anomaly.  We further confirm our results by showing that for the case in which the pump polarization is perpendicular to the applied magnetic field,  no  metastable response associated with chiral anomaly is observed.

\section*{Concept and Theory}
The chiral charge pumping process  leads to negative DC magneto-resistance (decrease of resistance in the presence of an applied parallel magnetic field) that is unusual among conventional metals/semimetals. The observation of DC  negative magneto-resistance in Weyl semiemetals was initially thought to be a conclusive experimental signature of the chiral anomaly  \cite{huangObservationChiralAnomalyInducedNegative2015c,zhangSignaturesAdlerBell2016c}.  However, it was later found that other classical effects  can give rise to similar negative DC magneto-resistance in semimetals, making it challenging for electrical measurements to discriminate the chiral  anomaly from other effects such as current jetting  \cite{arnoldNegativeMagnetoresistanceWelldefined2016b,liangExperimentalTestsChiral2018c}. The relaxation time of the chiral imbalance  in the Weyl semimetal TaAs was indirectly extracted  and estimated to be tens of picoseconds from negative magneto-resistance measurements \cite{zhangSignaturesAdlerBell2016c}. However, due to the large momentum separation between the Weyl nodes, the relaxation of unbalanced chiral Weyl carriers to their equilibrium state requires large momentum scattering and is theoretically expected to be much slower \cite{parameswaranProbingChiralAnomaly2014c}. 

\begin{figure}[htbp]
  \centering
  \includegraphics[scale=1]{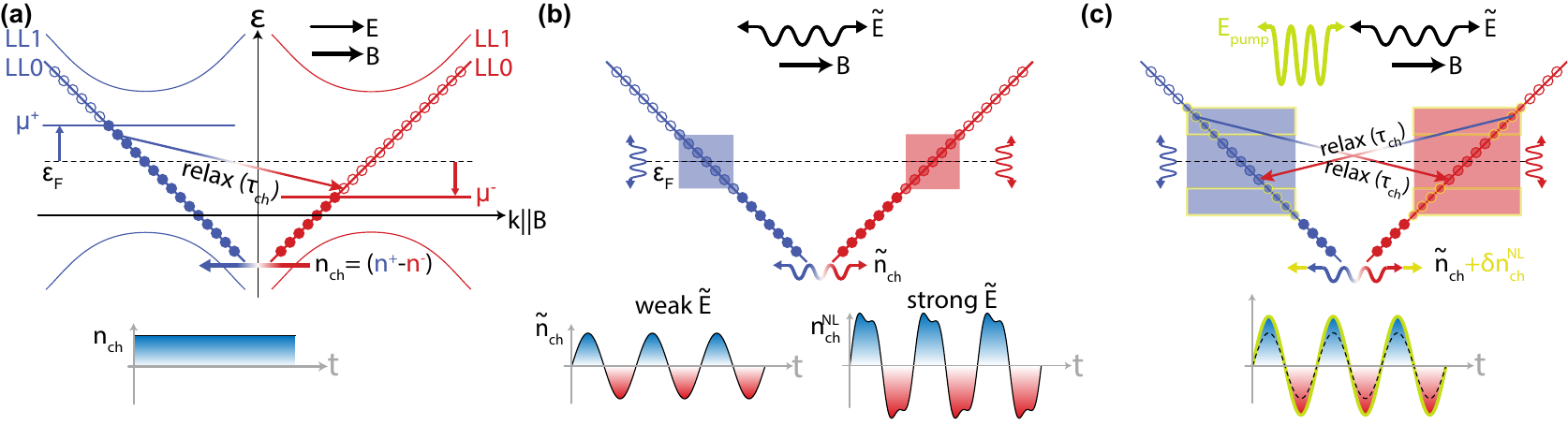}
  \caption{\textbf{Nonlinear Chiral Charge Pumping -- Concept and Experiment} (a) Static charge pumping of chiral carriers in the zeroth Landau level: applying parallel static electric $E$ and magnetic $B$ fields produces a chiral current $J_\textrm{ch}$ that pumps the carriers from one Weyl node to the other with opposite chirality and unbalances their chemical potential. Red and blue colors denote different chiralities of carriers. The chiral charge imbalance relaxes back to equilibrium with a time constant of $\tau_{ch}$.  (b) Dynamic charge pumping of chiral carriers in the zeroth Landau level: oscillating optical field   polarized parallel to a magnetic field $B$ causes harmonic oscillation of number of chiral carriers at the same frequency. This generates a distribution of chiral carriers around the equilibrium Fermi level $\epsilon_F$ shown by shaded blue and red regions.  For strong optical excitation, the chiral charge oscillation between the Weyl nodes becomes nonlinear in $\tilde{E}$ leading to anharmonic oscillation of chiral charges. (c) Nonlinear optical control of dynamic chiral anomaly: strong optical field $\tilde{E}_\textrm{pump}$ enhances the chiral current generated by $\tilde{E}||B$. When $\tilde{E}_\textrm{pump}$ is turned off, the extra pump-induced chiral carrier distribution (illustrated by a yellow border) slowly relaxes back via inter-Weyl-node relaxation. }
    \label{fig:1}
\end{figure}

We consider the extreme quantum limit where, upon the application of a magnetic field on  a Weyl semimetal, the only occupied Landau levels (LLs) are the $N=0$ chiral Landau levels (LL0s).  At equilibrium, the two LL0s for a pair of Weyl nodes in Weyl semimetals host equal number of left- and right-moving chiral carriers along the magnetic field. The static chiral charge pumping scheme is illustrated in Fig.~\ref{fig:1}a, where the application of parallel static electric $\mathbf{E}$ and magnetic $\mathbf{B}$ fields on a Weyl semimetal  unbalances the chemical potentials in the two LL0s with opposite chirality ($\mu^+$ and $\mu^-$). The chiral charge imbalance relaxes back to equilibrium with the chiral charge relaxation of $\tau_\textrm{ch}$. 

The chiral anomaly can also be realized in a dynamical  fashion  where, instead of constant chiral charge imbalance around the two Weyl nodes, their chemical potentials oscillate antisymmetrically around the equilibrium Fermi energy \cite{goswamiAxialAnomalyLongitudinal2015c,levyObservingOpticalSignaturesc}.  As illustrated in Fig.~\ref{fig:1}b, dynamical chiral anomaly can be achieved  by applying an oscillatory optical field of $\mathbf{E}(t)$ at frequency of $\omega$ parallel to a static $\mathbf{B}$ field.  In this case, the chiral carriers  oscillate in between the two nodes synchronously with $\mathbf{E}(t)$, leading to quasiparticle excitations near the chemical potential energy, illustrated by shaded blue and red regions in Fig.~\ref{fig:1}b. The density of the quasiparticles generated by  dynamical chiral charge pumping is $\tilde{n}_{\textrm{ch}}(\omega)/2$ in each Weyl node, where $\tilde{n}_{\textrm{ch}}(\omega)$ is the amplitude of the oscillatory chiral charge imbalance,

\begin{equation} \label{eq:1}
|\tilde{n}_{\textrm{ch}}(\omega)|=|n_{+}-n_{-}|=\frac{2e^2}{h^2}\frac{\tau_{ch}}{\sqrt{1+\omega^2\tau^2_{ch}}}\left|\mathbf{\tilde{E}}(\omega)\cdot\mathbf{B}\right|
\end{equation} 
$n_{+}$ and $n_{-}$ are densities of carriers with opposite chirality, and $\tilde{\mathbf{E}}(\omega)$ is the complex amplitude representing the magnitude and phase of the oscillating field. The excited quasiparticles relax back to the equilibrium chemical potential by  the time constant of $\tau_\textrm{ch}$. The associated magneto-optical conductivity to equation \eqref{eq:1} is (supplementary information),
\begin{equation} \label{eq:2}
\sigma_{\textrm{ch}}(\omega)=\frac{e^3v}{h^2}\frac{\tau_{ch}}{1-i\omega\tau_{ch}}B\approx i\frac{e^3v}{h^2\omega}B
\end{equation}  
where $v$ is  the Fermi velocity, and we have assumed $\omega\tau_\textrm{ch}\gg1$ in the second equality. We note that the  dynamical chiral anomaly scheme is only observable for excitations created by low-energy photons where the Weyl carriers with opposite chirality exists. For TaAs, this condition limits the photon energy to $\hbar \omega < 50$ meV (12 THz)  \cite{huangWeylFermionSemimetal2015c,arnoldChiralWeylPockets2016b}. In the limit of $\omega\rightarrow0$, \eqref{eq:1} and \eqref{eq:2} reproduce the  chiral and charge imbalance DC magneto-conductivity associated with the conventional static chiral anomaly \cite{sonChiralAnomalyClassical2013c}. 

We consider the case where $\tilde{\mathbf{E}}(\omega)$ is  sufficiently  intense  so that the resulting $\tilde{n}_{\textrm{ch}}$ becomes a  nonlinear function of the excitation field $\tilde{\mathbf{E}}(\omega)$, similar to anharmonic oscillation of an oscillator under a strong driving force. In this case,  $\tilde{n}_{\textrm{ch}}$ is not  simply  sinusoidal and can include harmonics of the driving frequency of $\omega$. This nonlinear scheme is illustrated in the lower right part of Fig.~\ref{fig:1}b.  We note that nonlinear optical response of Weyl semimetals driven  by Berry curvature of chiral carriers has been recently explored through measurements of second harmonic generation \cite{wuGiantAnisotropicNonlinear2017c} and the circular photogalvanic effect  \cite{maDirectOpticalDetection2017a,dejuanQuantizedCircularPhotogalvanic2017a,osterhoudt2019colossal,ma2019nonlinear}. In our study, we specifically  explore how a strong optical pump $\tilde{\mathbf{E}}_{\textrm{pump}}$ can modify and control the chiral anomaly conductivity by driving it to the nonlinear regime. As illustrated  in Fig.~\ref{fig:1}c, the strong optical pump $\tilde{\mathbf{E}}_{\textrm{pump}}$ enhances the quasiparticle density produced by dynamical chiral charge pumping  in equation \eqref{eq:1}.  The enhancement is calculated as (supplementary information),

\begin{equation} \label{eq:3}
\delta\tilde{n}_{\textrm{ch}}(\omega)=\frac{15\alpha^2e^4v^2}{4h^2\omega^4\tau}\left(\frac{\mathbf{\tilde{E}}_{\textrm{pump}}\cdot\mathbf{B}}{B}\right)^2 \mathbf{\tilde{E}}\cdot\mathbf{B},
\end{equation}  
where $\alpha \equiv 2/mv^2$ and $m$ is the mass associated with the finite Lifshitz transition energy of the Weyl bands which is the origin of the nonlinearity. 

The nonlinear contribution $\delta\sigma_{\textrm{ch}}^{\textrm{NL}}$ to the chiral anomaly conductivity can be written as $\sigma_{\textrm{ch}}\left(\omega,\tilde{E}_\textrm{pump}(\omega)\right)=\sigma_{\textrm{ch}}(\omega)+\delta\sigma_{\textrm{ch}}^{\textrm{NL}}$,  where 

  \begin{equation} \label{eq:4}
\delta\sigma_{\textrm{ch}}^{\textrm{NL}}=i\frac{9\alpha^2e^5v^3}{8h^2\omega^3}\left(\frac{\mathbf{\tilde{E}}_{\textrm{pump}}\cdot\mathbf{B}}{B}\right)^2 B,
\end{equation} 
 which exhibits a linear dependence on $B$ similar to the linear conductivity in \eqref{eq:2}. 

The optical control of chiral anomaly in dynamical fashion is demonstrated in Fig.~\ref{fig:1}c. The nonlinear process described by equations \eqref{eq:3} and \eqref{eq:4} predicts that the chiral current can be optically enhanced. Furthermore, by driving the chiral anomaly to the nonlinear regime and observing the relaxation dynamics in the time domain, one can measure the chiral charge relaxation with time constant $\tau_\textrm{ch}$, i.e. how long it takes for the excited carriers in the chiral Landau level to relax back to the equilibrium chemical potential. Since $\tau_\textrm{ch}$ is expected to be long, the optical modification of chiral current will be long-lived. As shown in the supplementary information, assuming parameters for TaAs and optical pumping with field strength of 50 kV/cm (achievable in our pulsed THz pumping experiments), $\delta\sigma_{\textrm{ch}}^{\textrm{NL}}/\sigma_{\textrm{ch}}\approx1$, suggesting   the optical pumping can fully modify the chiral charge pumping process. We note that the long-lived optical control of chiral charge pumping described by equations \eqref{eq:3} and \eqref{eq:4} is relevant only in the quantum limit where the carriers occupy only the zeroth Landau level LL0. In the semiclassical regime, where many Landau levels are occupied, the pump-excited chiral carriers can relax back to equilibrium chemical potential by inter-Landau level scattering within each Weyl node. 

 \section*{Weyl Semimetal Sample}
The Weyl semimetal considered in our study is tantalum arsenide (TaAs). Single crystals of TaAs were grown using the chemical vapor transport method with iodine as the transport agent \cite{osterhoudt2019colossal}. Unlike the other discovered Weyl semiemtals, TaAs has a Fermi energy very close to the Weyl nodes in the linear regime of the bands, so that the electrons at the  Fermi surface behave like massless chiral Weyl fermions \cite{huangWeylFermionSemimetal2015c,xuDiscoveryWeylFermion2015c,lvExperimentalDiscoveryWeyl2015c,arnoldChiralWeylPockets2016b}. This makes TaAs especially attractive for studies of Weyl fermions interactions with light. TaAs has two types of Weyl nodes namely W1 (4 pairs) and W2 (8 pairs), where W2 nodes are closer to the chemical potential than W1  by about 12 meV \cite{arnoldChiralWeylPockets2016b}. We estimate that for $B  \ge 2$ T, the W2 Weyl nodes are in the extreme quantum limit, whereas W1 carriers are always in the semiclassical regime occupying few Landau levels (supplementary information). In tantalum arsenide, for ($\gtrsim$ 50 meV) energies, the pairs of Weyl bands merge into single bands, causing Weyl bands and Weyl fermions to exist only at low energy excitation \cite{huangWeylFermionSemimetal2015c,arnoldChiralWeylPockets2016b}. Also, other than W1 and W2 Weyl bands, the chemical potential in TaAs crosses a non-Weyl band with a band gap of $\lesssim$ 50 meV  \cite{huangWeylFermionSemimetal2015c,lee2015fermi,arnoldChiralWeylPockets2016b}. Therefore, in order to probe the dynamics of Weyl fermions and not to excite the other non-Weyl carriers in TaAs, photon energy below 50 meV is required.

\section*{Experimental Setup}

\begin{figure}[htbp]
  \centering
  \includegraphics[scale=1.5]{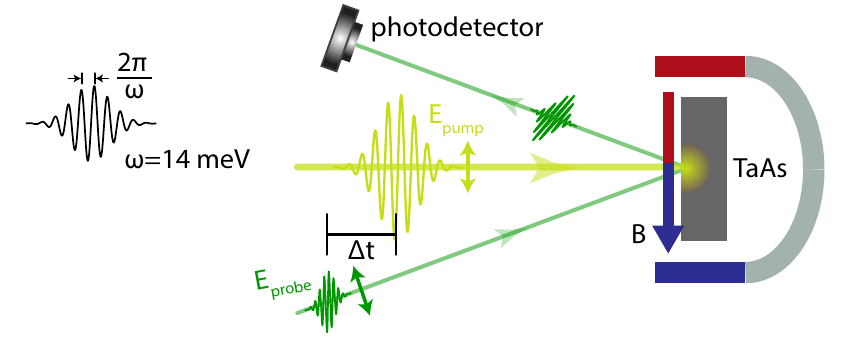}
  \caption{\textbf{Experimental Setup}The pump-probe setup at the photon energy of $\hbar \omega$=14 meV in  reflection geometry. The pump and probe pulses are co-polarized to the (100) face of TaAs crystal that is placed inside a magnet with B field parallel to pump/probe polarization. }
    \label{fig:12}
\end{figure}

In order to experimentally test the idea of nonlinear optical control of chiral anomaly and measure the chiral population  relaxation time $\tau_{\textrm{ch}}$ in a Weyl semimetal, we design pump-probe experiments at low photon energy (terahertz)  in a magnetic field on TaAs. All the pump-probe measurements were carried out using quasi-continuous wave pulses at a photon energy of 14 meV ($\equiv 3.4$ THz), which is well inside the Weyl bands where Weyl fermions with opposite chirality exist \cite{arnoldChiralWeylPockets2016b}. Furthermore, at 14 meV, the photons do not have enough energy to excite additional carriers in the non-topological bands \cite{huangWeylFermionSemimetal2015c}. The experimental pump-probe setup is illustrated in Fig.~\ref{fig:12}. The TaAs crystal is mounted inside a magnet with a  magnetic field parallel to the ab plane of the crystal. An intense pump pulse polarized parallel to  applied  magnetic field ($\mathbf{E}_{\textrm{pump}} \parallel \mathbf{B}$)  drives the chiral charge pumping process to the nonlinear regime. In order to monitor the pump-induced  change in the conductivity, reflection of a weak probe pulse ($\mathbf{E}_{\textrm{probe}} \parallel \mathbf{B}$) is measured as a function of the time delay between the pump and probe pulses.

Equations \eqref{eq:3} and \eqref{eq:4} predict pump-induced increase in the quasiparticle density and conductivity. Therefore, the signature of pump-induced nonlinearity in the chiral anomaly is a long-lived positive change in the probe reflection (details in the supplementary information section IV). We note that  other nonlinearities are present, such as hot carriers effects, that can contribute to a pump-induced change in the probe reflection. To distinguish those, we carry out measurements at different magnetic fields, pump fluences, and pump polarizations. Furthermore, since the  nonlinearities have different time scales, the time delay scan helps us differentiate various pump-induced nonlinearities.  
\section*{Zero Magnetic Field}
\begin{figure}[htbp]
  \centering
  \includegraphics[scale=1.5]{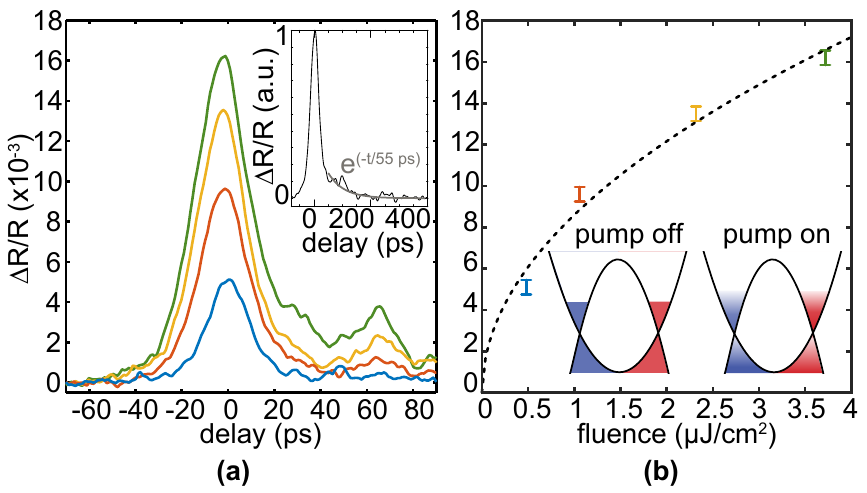}
  \caption{\textbf{Pump-probe at 14 meV, $\boldsymbol{B}$=0} (a) Fractional pump-induced increase in probe reflection for variety of pump fluences as a function of pump-probe time delay. Pump-probe traces exhibit two relaxation time constants of $<$15 ps (pulsewidth-limited) and $\approx$55 ps, as shown in the inset.  (b) Maximum change (occurs at zero time delay) in probe reflection as a function of pump fluence extracted from part (a). The dashed curve is a power-law fit exhibiting square-root dependence of the peak value on the pump fluence, consistent with the pump-induced hot carriers in the Weyl bands.  
 }
  \label{fig:2}
\end{figure}

Fig.~\ref{fig:2} shows the pump-probe measurement results without an applied magnetic field. In Fig.~\ref{fig:2}a, the measured relative pump-induced change in the probe reflection is plotted as a function of pump-probe time delay for variety of pump fluences. We observe a positive change in probe reflection that increases with the pump fluence. The  probe reflection relaxes immediately back to its equilibrium  level through a fast process that cannot be resolved by the pulsewidth used here followed by a slower relaxation tail. The slower relaxation process can be fit to an exponential function with a time constant of 55 ps, as shown by the grey curve in the inset of Fig.~\ref{fig:2}a.

The zero magnetic field results in Fig.~\ref{fig:2}a can be understood by the effect of carrier heating in the Weyl bands of TaAs. The pump pulse heats up the carriers in the Weyl bands to a temperature $T$ higher than the initial lattice temperature (10 K). This process leads to the excitation of hot carriers around the Weyl nodes, which elevates the Drude weight by an amount proportional to $T^2$ in a 3D Dirac/Weyl semimetal \cite{hosurChargeTransportWeyl2012c}. The reflection, which is linearly proportional to the Drude weight, will also increase by  $\Delta R \propto T^2$. We note that the quadratic increase of reflection with temperature has been previously observed in TaAs \cite{xuOpticalSpectroscopyWeyl2016c} and other 3D Dirac/Weyl semimetals \cite{jenkinsThreedimensionalDiracCone2016d}. From the temperature dependence of the linear reflection data in TaAs \cite{xuOpticalSpectroscopyWeyl2016c}, we estimate the pump-induced carrier temperature rise to be around 50 K for the highest pump fluence considered here. 
After the pump pulse, the carriers start to cool down via phonon emissions, and the increased reflection relaxes back to the equilibrium value.  Thus, the two relaxation time constants of $\ll$30 ps and 55 ps are related to the electron-phonon relaxation times. Since both the photon energy (14 meV) and the thermal energy associated with the carrier temperature rise are below the energy of optical phonons in TaAs, the hot carrier relaxation will occur through acoustic phonons  \cite{lundgrenElectronicCoolingWeyl2015d}. A possible scenario for electron-phonon relaxation  observed in our experiments is that the faster process is a disorder-assisted electron-phonon relaxation process, whereas the slower one is related to conventional electron-acoustic phonon collisions \cite{lundgrenElectronicCoolingWeyl2015d}. We note that in graphene, as another Dirac semimetal, similar processes govern the hot carrier relaxations  \cite{song2012disorder, jadidi2016tunable} and pump-probe dynamics \cite{graham2013photocurrent,jadidi2016nonlinear}. For disorder-assisted electron-phonon cooling in 3D Dirac/Weyl semimetals, the carrier temperature rise is $T \propto F^{1/4}$, where $F$ is the pump fluence \cite{lundgrenElectronicCoolingWeyl2015d}. Since  $\Delta R \propto T^2$, we have $\Delta R \propto \sqrt{F}$. Therefore, close to the zero time delay where the disorder-assisted process governs the relaxation, we expect square-root dependence of $\Delta R/R$ on the pump fluence $F$.  This is demonstrated in Fig.~\ref{fig:2}b where we plot $\Delta R/R$ around zero time delay as a function of the pump fluence. The dashed curve is a square root fit exhibiting an excellent match to the data. 

\section*{Magnetic Field Dependence}
\begin{figure}[htbp]
  \centering
  \includegraphics[scale=1]{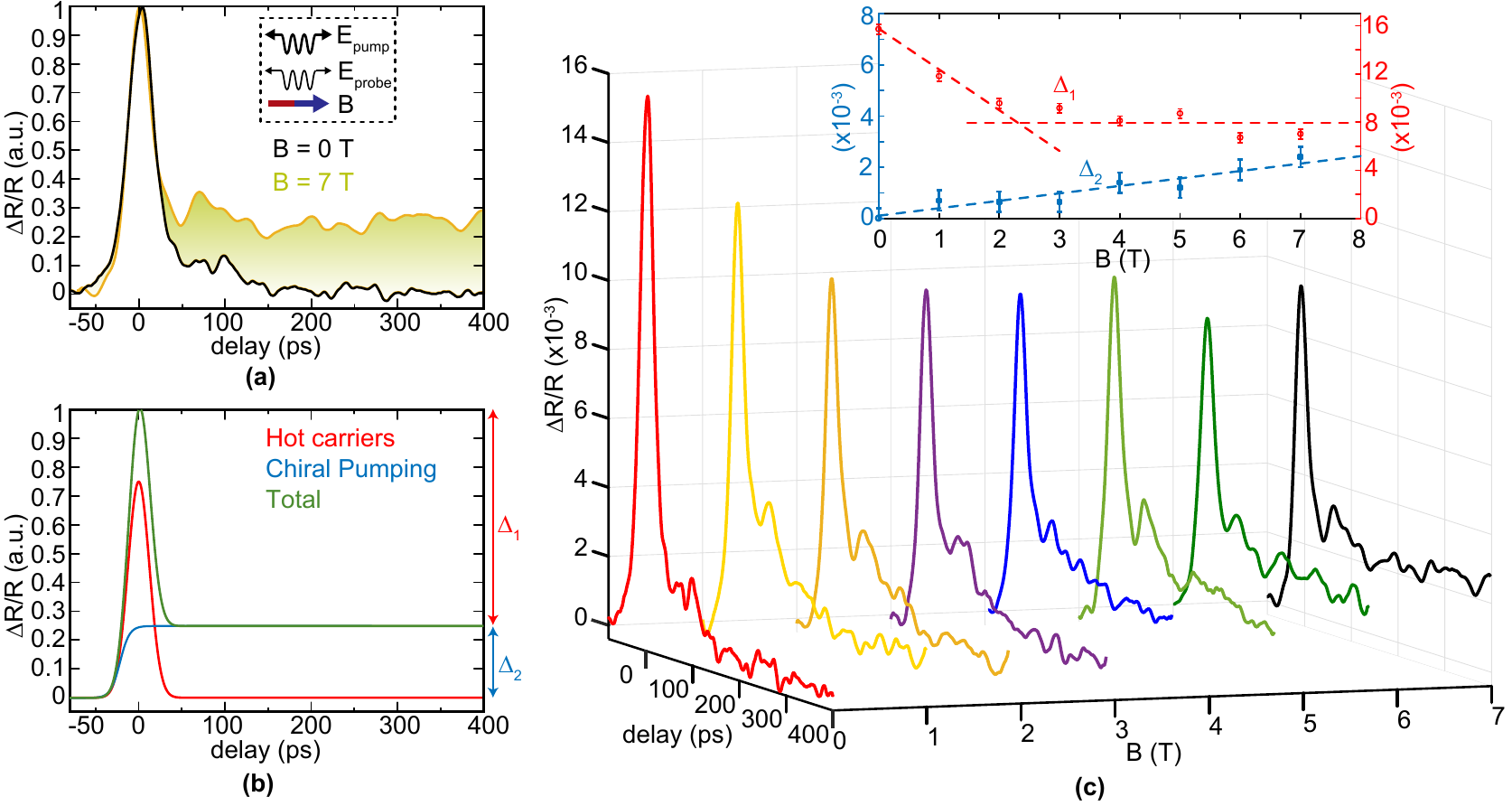}
  \caption{$\boldsymbol{E_\textrm{pump}\parallel E_\textrm{probe} \parallel B }$ \textbf{(varying $\boldsymbol{B}$).}  (a) Pump-induced increase in probe reflection for $B$=0 T (black) and $B$=7 T (yellow) in arbitrary units as a function of pump-probe time delay. The pump-probe trace at $B$= 7 T exhibits a very slow response ($\gg$1 ns).  The peak around 60 ps is caused by a replica-- the reflected pump pulse from the cryostat window.  (b)  Expected pump-probe traces for fast hot carriers effects (red), metastable chiral pumping (blue), and the net result (green). $\Delta_1$ is defined as the maximum pump-induced change in probe reflection which happens around zero time delay, and $\Delta_2$ characterizes long-lasting reflection changes that does not relax to equilibrium in the considered time delay range. (c)  Pump-probe trace for different static magnetic fields applied parallel to pump/probe polarization. The inset shows $\Delta_1$ (red) and $\Delta_2$ (blue) as a function of the applied magnetic field. 
  }
  \label{fig:3}
\end{figure}

We now discuss the pump-probe measurements in the presence of an applied static magnetic field. In Fig.~\ref{fig:3}a, we plot the normalized pump-induced relative change in probe reflection for $B$=0 T (black) and $B$=7 T (yellow). The two curves exhibit similar pump-probe responses close to the zero time delay, suggesting that the initial response for $B$=7 T is also related to pump-induced hot carriers. At longer time delays, when the carriers are expected to be cooled down to the quasi-equilibrium temperature, the pump-probe trace at $B$= 7 T exhibits a metastable positive response that does not fully recover within the 400 ps measurement range considered here. This suggests a process with a time constant of much greater than 1 ns but smaller than the repetition period of  pulses  (77 ns). The positive sign of the metastable signal suggests a  pump-induced positive change in the conductivity, consistent with the nonlinear chiral charge pumping  (equation \eqref{eq:3}).  Based on this picture, we interpret the metastable process to be the chiral pumping relaxation time $\tau_{ch}$ which we estimate to be  (1 ns $\ll \tau_{ch} <$ 77 ns). This occurs in the $N=0$ chiral Landau level, which allows long-lived excitations due to the reduced phase space for scattering. As presented later in this manuscript, the metastable  response disappears when $\mathbf{E}$ and $\mathbf{B}$ are perpendicular to one another. In Fig.~\ref{fig:3}b, we illustrate the expected pump-probe traces for fast hot carriers response (red), metastable chiral charge pumping response (blue), and the net response (green).  As depicted in the right side of plot in Fig.~\ref{fig:3}b, we define $\Delta_1$ as the size of the pump-probe signal at zero time delay, and $\Delta_2$ as the size of the ultraslow signal. The long relaxation time of chiral charge pumping agrees well with theoretical estimations and is expected, given the large momentum transfer needed for carriers to transport between the Weyl nodes \cite{parameswaranProbingChiralAnomaly2014c}. We note that the measured $\tau_{ch}$ is much larger than the values inferred indirectly from the negative magnetoresistance measurements \cite{zhangSignaturesAdlerBell2016c}, although there reamins  uncertainty about whether the negative magnetoresistance measurements are entirely attributed to the chiral anomaly \cite{liangExperimentalTestsChiral2018c}. The observation of terahertz-excited  $\omega=(2\pi)3.4\times10^{12}$ rad/s chiral anomaly with a long relaxation time ($\omega\tau_\textrm{ch}>10^4$)   holds promises for applications of Weyl semiemtals in quantum technology \cite{kharzeev2019}. In order to further explore the observed metastable process and its relation to chiral charge pumping, we investigate the dependence of the metastable signal on the applied static magnetic field and the pump fluence. 

In Fig.~\ref{fig:3}c, we show the measured pump-probe traces  for variety of applied static magnetic fields from $B$=0 T to 7 T. The metastable response is present at all nonzero $B$ values. As we increase $B$, the peak of the pump-probe response at zero time delay shrinks, while the level of the  ultraslow signal  increases. We extract $\Delta_1$ and $\Delta_2$  from the data in Fig.~\ref{fig:3}b and plot them as a function of the applied magnetic field in the inset. This figure clearly demonstrates the decrease (increase) of $\Delta_1$ ($\Delta_2$) with  magnetic field. We note that this behavior is consistent with the origin of $\Delta_1$ and $\Delta_2$ to be related to hot carriers and nonlinear chiral charge pumping respectively: As we increase the applied magnetic field from zero, carriers in the W2 Weyl node go to the  quantum limit, so that for $B \approx 2$ T, all W2 carriers are in LL0 (supplementary information). 
 Since the Drude weight of carriers in LL0 is independent of temperature \cite{goothExperimentalSignaturesMixed2017a}, the W2 carriers contribution to the initial hot carriers  response diminishes as $B$ is increased, and hence the decrease of $\Delta_1$. As for $\Delta_2(B)$, according to \eqref{eq:3}, the pump-induced nonlinear magneto-optical conductivity increases with $B$ causing the reflection change to increase. This predicts a linear dependence of $\Delta_2$ on $B$, which is consistent with the experimental observations, as shown by the dashed blue curve in Fig.~\ref{fig:3}c. In the supplementary information, we carefully examine the contribution of carriers in different Weyl bands of W1 and W2  and the trivial carriers in TaAs to the observed pump-probe signals at zero and finite magnetic fields.

\section*{Pump Fluence Dependence}
\begin{figure}[htbp]
  \centering
  \includegraphics[scale=1.5]{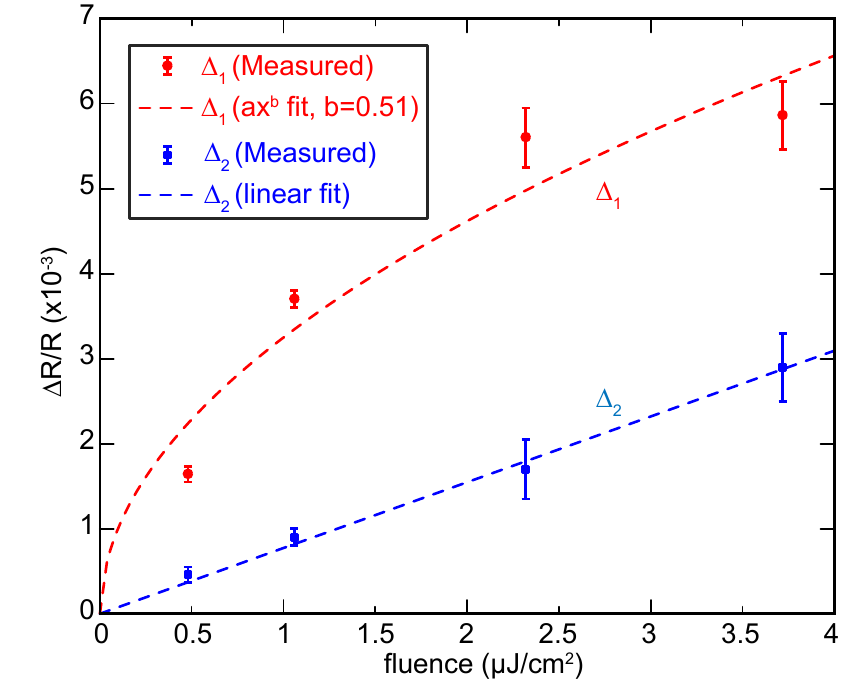}
  \caption{$\boldsymbol{E_\textrm{pump}\parallel E_\textrm{probe} \parallel B }$ \textbf{(=7 T).} $\Delta_1$ (red) and $\Delta_2$ (blue) as a function of pump fluence. The red (blue) dashed lines are square-root (linear) fit to $\Delta_1$ ($\Delta_2$) respectively. 
}
  \label{fig:4}
\end{figure}

We investigate the dependence of $\Delta_1$ (peak of the pump-probe signal) and $\Delta_2$ (ultraslow nonlinear chiral anomaly signal) at $B=7$ T on the pump fluence. As shown in Fig.~\ref{fig:4}, $\Delta_1$ is found to exhibit a sublinear dependence on the pump fluence. A power-law fit to the measured $\Delta_1$ vs pump fluence reveals a square-root-like dependence (dashed red curve in Fig.~\ref{fig:4}), similar to the peak-vs-fluence dependence observed at $B=0$ T (Fig.~\ref{fig:2}b). The similarity to $B=0$ T measurements again points toward analogous hot carrier  origin of $\Delta_1$ at nonzero magnetic fields. On the other hand, as seen in Fig.~\ref{fig:4}, $\Delta_2$ exhibits a perfectly linear dependence of the pump fluence. This suggests that the conductivity change is linear in the pump fluence and is in excellent agreement with the pump-induced change in the chiral pumping  conductivity calculated in equation \eqref{eq:3}.

\section*{Polarization Dependence}
\begin{figure}[htbp]
  \centering
  \includegraphics[scale=1.5]{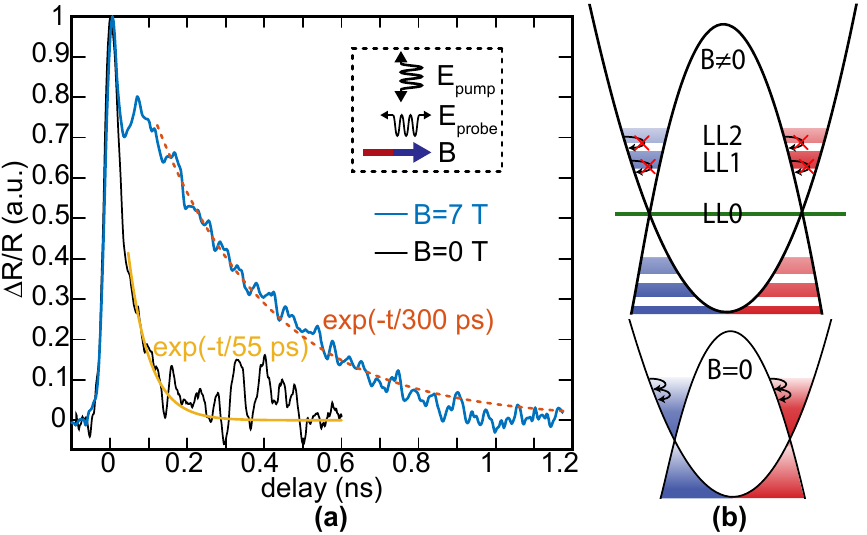}
  \caption{$\boldsymbol{E_\textrm{pump}\perp E_\textrm{probe} \parallel B }$ \textbf{(=7 T).} (a) Relative pump-induced increase in probe reflection for variety of pump fluences as a function of pump-probe time delay. (b) Carriers in Landau levels (LLs) of a Weyl semimetal. The phase space for phonon scattering of hot carriers in LLs (top panel) is strongly suppressed compared to zero magnetic field (bottom panel). 
%
}
  \label{fig:5}
\end{figure}

To further confirm our interpretation of our results, we present the pump-probe measurement results with pump polarized perpendicular to the applied magnetic field, while keeping probe polarized parallel to the field.  In this case, we do not expect the pump pulse to excite a dynamical chiral current or change the chiral pumping conductivity. In Fig.~\ref{fig:5}a, the blue curve is the  measured pump-probe response at $B=7$ T under these conditions. First we note that the ultraslow plateau observed for $\mathbf{E}_{\textrm{pump}} \parallel \mathbf{B}$ is absent, supporting the nonlinear chiral charge pumping origin of the ultraslow signal.
Comparing the $B=7$ T result to $B=0$ T in  Fig.~\ref{fig:5}a, two main differences are noted: (i) the  fast pulsewidth-limited relaxation process around the zero time delay is no longer present, and (ii) instead of the 55 ps tail, a  slower process with the time constant of 288 ps extracted from the exponential fit (red dashed curve) appears. According to the hot carriers nature of the $B=0$ T results described earlier, the two differences suggest that the pump-induced hot carriers relaxation processes are strongly suppressed in the presence of an applied magnetic field perpendicular to the pump polarization. This is expected due to the strong reduction of phase space for carriers phonon scattering in highly quantized Landau levels to scatter off the phonons, as illustrated in Fig.~\ref{fig:5}b.  Similar effects have been reported in the from pump-probe measurements on graphene in a magnetic field \cite{mittendorffIntrabandCarrierDynamics2014c}. Therefore, as expected from equation \eqref{eq:3}  the observed metastable process for $\mathbf{E}_{\textrm{pump}} \parallel \mathbf{B}$ attributed to pump-induced chiral charge pumping is absent for the case of $\mathbf{E}_{\textrm{pump}} \perp \mathbf{B}$. In the supplementary information, we present more pump-probe data for $\mathbf{E}_{\textrm{pump}} \perp \mathbf{B}$ case at different magnetic fields and pump fluences that all exhibit time dynamics similar to $B=7$ T shown in Fig.~\ref{fig:5}a.

\section*{Conclusion}
Our experimental and theoretical results confirm our proposed on a nonlinear optical scheme to manipulate the chiral charge pumping which is a unique  transport property of topologically-protected Weyl fermions. Using time-resolved optical techniques, we measure the relaxation of chiral anomaly in the Weyl semimetal TaAs to be very long ($\tau_\textrm{ch}\gg 1$ ns). Our findings pave the way for future studies  exploring optical control of topological transport phenomena in solids, and investigating long-lived chiral Weyl fermions for quantum optoelectronic applications.


\section*{Methods}
\textbf{Terahertz Pump-probe Measurements in a Magnetic Field:} 
 The free electron laser was tuned to produce 30 ps wide pulses at
a  photon energy of 14 meV and a repetition rate of 13 MHz. The beam was split into pump and probe beams that were delayed relative to one another using a mechanical delayline. The probe pulse is polarized along the applied static magnetic field that is oriented parallel to the a-axis of the TaAs crystal. The pump pulse, which is in all cases more than 20 times stronger than the probe pulse,  is co-polarized with the probe pulse, except for the measurements presented in Fig.~\ref{fig:5}. The pump and probe beams were overlapped and focused using an off-axis parabolic mirror onto the TaAs crystal located inside a magnet (Voigt geometry). The magnetic field can be tuned from 0 T to 7 T. The TaAs crystal was cooled to a temperature of 10 K for all of the pump-probe measurements. The emerging pump beam was extinguished while the reflected probe beam was measured using a cryogenically cooled bolometer as a function of the pump-probe delay $\Delta t$.
\\

\section*{Acknowledgements}
Work at UCLA was supported by the U.S. Department of Energy (DOE), Office of Science, Office of Basic Energy Sciences under Award Number DE-SC0011978. M. K. acknowledges the support from the Sharif University of Technology under Grant No. G690208.


\bibliographystyle{naturemag}
\bibliography{REF3}

\pagebreak
\part*{Supplementary Information}

\renewcommand{\theequation}{S\arabic{equation}}
\renewcommand{\thesection}{S\arabic{section}}
\renewcommand{\thefigure}{S\arabic{figure}}
\setcounter{equation}{0}
\setcounter{section}{0}
\setcounter{figure}{0}

%
%

\def\be{\begin{equation}}
\def\ee{\end{equation}}
\def\lra{\longrightarrow}
\def\bi{\begin{itemize}}
\def\ei{\end{itemize}}
\def\bn{\begin{enumerate}}
\def\en{\end{enumerate}}
\def\bea{\begin{eqnarray}}
\def\eea{\end{eqnarray}}
\newcommand{\bpm}{\begin{pmatrix}}
\newcommand{\epm}{\end{pmatrix}}
\def\no{\nonumber}
\def\ba{\begin{array}}
\def\ea{\end{array}}
\def\bd{\begin{displaymath}}
\def\ed{\end{displaymath}}
\def\la{\langle}
\def\ra{\rangle}
\renewcommand{\imath}{\hspace{1pt}\mathrm{i}\hspace{1pt}}

\renewcommand{\vec}{\mathbf}
\renewcommand{\Re}{\mathop{\mathrm{Re}}\nolimits}
\renewcommand{\Im}{\mathop{\mathrm{Im}}\nolimits}


\vspace{1cm}
This supplemental materials contain the details of the  theory we developed to understand the main findings of our optical measurements. We begin by a brief introduction of Weyl nodes appearing in band structure of solids in Sec. \ref{TWS}. Then, since we are interested in the optical transport of the system in the extreme quantum limit, we discuss the transport of the extreme quantum limit in Sec. \ref{strong_B}, where only zeroth Landau level (LL0) is occupied. Then, a lattice model for Weyl semimetals and the formation of LLs are presented in Sec. \ref{lattice}.  Details of the pump-probe signal from TaAs alongside with the calculations of the reflection are given in Sec. \ref{PP_signal}.

\section{Wely Semimetals: general remarks \label{TWS} }
We consider a doped Weyl semimetal whose low-energy excitations near the nodes are described by the following Hamiltonian:
\bea H_{\eta}=\eta \hbar v \boldsymbol{\sigma}\cdot\mathbf{k} -\varepsilon_{F}, \eea
where the wave vector $\mathbf{k}=(k_{x},k_{y},k_{z})$ is measured from the node and $\boldsymbol{\sigma}$ is a vector of Pauli matrices. Here, $v$ is velocity and $\varepsilon_{F}$ is the Fermi energy. For simplicity we consider the system is composed of two Weyl nodes with opposite chiralities $\eta=\pm$. In the momentum space the Weyl nodes act as source and sinks of Berry curvature $\boldsymbol{\Omega}^{\eta}_{\mathbf{k}}=\boldsymbol{\nabla}\times \mathbf{A}^{\eta}_{\mathbf{k}}$, where $\mathbf{A}^{\eta}_{\mathbf{k}}=i\langle u^{\eta}_{\mathbf{k}}|\boldsymbol{\nabla}u^{\eta}_{\mathbf{k}}\rangle$ is the Berry connection of a Bloch wave function $u^{\eta}_{\mathbf{k}}$ of a given node and $H_{\eta}u^{\eta}_{\mathbf{k}}=\varepsilon_{\mathbf{k}}u^{\eta}_{\mathbf{k}}$. It yields $\boldsymbol{\Omega}^{\eta}_{\mathbf{k}}=\eta\hat{\mathbf{k}}/2k^2$. The chirality of of given node is given by the total flux of the Berry curvature as $\eta=1/2\pi \oint \boldsymbol{\Omega}^{\eta}_{\mathbf{k}}\cdot d\mathbf{S}_{k}$, where the integral is taken over a closed surface enclosing the nodes in the momentum space.

\section{Optical effects in the extreme quantum limit \label{strong_B}} 
 The Weyl cones are replaced by dispersive Landau levels in a strong magnetic field. The carriers are transported through the LLs, indexed by $n$, in an electric field directed along the $z$ axis. The transport is described by the linearized Boltzmann kinetic equation for the node with chirality $\eta$ as 
  
\bea \label{Blz} \partial_{t}f_{n}^{\eta}(p_{z})+eE\partial_{p_{z}} f_{n}^{\eta}(p_{z})=I\{f_{n}^{\eta}(p_{z})\}, \eea  where $f_{n}^{\eta}(p_{z})$ is the distribution function for electron states with momentum $p_{z}$ in the $n$-th LL and we take $e=-|e|$. Here $I$ on the right-hand side denotes the collision integrals. We assume that the momentum relaxation rate within each valley $\tau^{-1}_{intra}$ is much larger than inter-valley relaxation rate $\tau^{-1}$, i.e., $\tau\gg \tau_{intra}$. Thus, the distribution function $f_{n}^{\eta}(p_{z})$ becomes isotropic in momentum and depends on energy as $f_{n}^{\eta}(p_{z})=f^{\eta}(\varepsilon_{n}(p_{z}))$, where $\varepsilon_{n}(p_{z})=\pm v\sqrt{2n\hbar|e|B+p_{z}^2 }$ for $n=1,2,\cdots$ and $\varepsilon_{0}(p_{z})=-\eta vp_{z}$ for $n=0$. Therefore \eqref{Blz} can be rewritten as 
 \bea \partial_{t}f^{\eta}(\varepsilon_{n}(p_{z}))+eE\partial_{p_{z}} f^{\eta}(\varepsilon_{n}(p_{z}))=I\{f^{\eta}(\varepsilon_{n}(p_{z}))\}. \eea  

Multiplying by $\sum_{n}\delta(\varepsilon-\varepsilon_{n}(p_{z}))$ and integrating over momentum, the above Boltzmann equation can be cast in the form

\bea \partial_{t}f^{\eta}(\varepsilon)+\frac{\eta}{\nu^{\eta}(\varepsilon)}\frac{e^2 \mathbf{E}\cdot\mathbf{B}}{h^2c} \partial_{\varepsilon}f^{\eta}(\varepsilon)= I\{f^{\eta}(\varepsilon)\}, \eea where
\bea \nu^{\eta}(\varepsilon)=\frac{1}{2\pi l_{B}^2}\sum_{n}\int \frac{dp_{z}}{h}\delta(\varepsilon-\varepsilon_{n}(p_{z}))=\frac{1}{2\pi l_{B}^2}\frac{1}{hv}\left(1+2\sum_{n=1}\frac{\epsilon}{\sqrt{\epsilon^2-2n}}\right) \eea
is the density of states with $l_{B}=\sqrt{\hbar /|e|B}$ as the magnetic length and $\epsilon=\varepsilon/\hbar v l_{B}^{-1}$. We use the relaxation time approximation and write the collision integral as 
 
\bea \label{Boltz_LL} \partial_{t}f+\frac{\eta}{\nu(\varepsilon)}\frac{e^2 \mathbf{E}\cdot\mathbf{B}}{h^2c} \partial_{\varepsilon}f= -\frac{f-f_{0}}{\tau(\varepsilon)}. \eea

To study the response of the system, we assume a monochromatic incident electric field with frequency $\omega$, i.e. $\mathbf{E}(t)=\mathbf{E} e^{-i\omega t}+\mathbf{E}^* e^{i\omega t}$, disturbs the distribution function as
\bea f=\sum_{n=0}^{\infty} f_{n}e^{-in\omega t},\eea where the terms $f_{n}$'s ($n\ge1$) are induced by the electric field. We truncate the series up to second order as \cite{morimotoSemiclassicalTheoryNonlinear2016c}  
\bea f=f_{0}+f_{1}e^{-i\omega t}+f_{2}e^{-2i\omega t}.\eea

The \eqref{Boltz_LL}  then becomes 
\bea \left[-i\omega f_{1}e^{-i\omega t}-2i\omega f_{2}e^{-2i\omega t}\right]+\frac{\eta}{\nu(\varepsilon)}\frac{e^2 \mathbf{E}(t)\cdot\mathbf{B}}{h^2c}\left[ \partial_{\varepsilon}f_{0} + \partial_{\varepsilon}f_{1} e^{-i\omega t}+ \partial_{\varepsilon}f_{2} e^{-2i\omega t} \right]\nonumber\\=-\frac{1}{\tau(\varepsilon)}\left[ f_{1}e^{-i\omega t}+f_{2}e^{-2i\omega t} \right]. \eea

To measure the response to the probe field we consider a configuration for the electric fields as $\mathbf{E}=\mathbf{E}_\textrm{pu}+\mathbf{E}_\textrm{pr}$, where the lowercase indices denote the pump and probe components, respectively. We calculate the response linear in $\mathbf{E}_\textrm{pr}$, while the amplitude itself might have been modulated by the pump field $\mathbf{E}_\textrm{pr}$ yielding a non-linear signal as discussed below. Plugging the electric field $\mathbf{E}(t)$ in the above equation and equating the terms proportional to $e^{-i\omega t}$ and $e^{-2i\omega t}$, we obtain the following expressions for $f_1=f_{1}^{(1)}+f_{1}^{(3)}$, where linear and non-linear terms read as 

\bea f_{1}^{(1)}=-\frac{\tau(\varepsilon)}{1-i\omega\tau(\varepsilon)}\frac{\eta}{\nu(\varepsilon)}\frac{e^2 \mathbf{E}_\textrm{pr}\cdot\mathbf{B}}{h^2} \partial_{\varepsilon}f_{0} \eea and 
\bea f_{1}^{(3)}=-\frac{3e^2 \mathbf{E}_\textrm{pr}\cdot\mathbf{B}}{h^2} \left(\frac{e^2 \mathbf{E}_\textrm{pu}\cdot\mathbf{B}}{h^2}\right)^2  \frac{\tau(\varepsilon)}{1-i\omega\tau(\varepsilon)} \frac{\eta}{\nu(\varepsilon)} \partial_{\varepsilon} \left( \frac{\tau(\varepsilon)}{1-2i\omega\tau(\varepsilon)} \frac{1}{\nu(\varepsilon)} \partial_{\varepsilon} \left( \frac{\tau(\varepsilon)}{1-i\omega\tau(\varepsilon)} \frac{1}{\nu(\varepsilon)} \partial_{\varepsilon}f_{0} \right) \right).\nonumber \\ \eea

\subsection{Linear magneto-conductivity}
At zero temperature the linear term $f_{1}^{(1)}$ gives rise to the chiral charge density at the node $\eta$   

\bea n_{\eta}=\int d\varepsilon~ \nu(\varepsilon) f^{(1)}_{1}(\varepsilon)=\frac{\tau_\textrm{ch}}{1-i\omega\tau_\textrm{ch}}\frac{\eta e^2 \tilde{\mathbf{E}}_\textrm{pr}\cdot\mathbf{B}}{h^2}, \eea
where $\tau_\textrm{ch}=\tau(\varepsilon_F)$. It then follows that

\bea \tilde{n}_\textrm{ch}(\omega)=n_{+}-n_{-}=\frac{2e^2}{h^2}\frac{\tau_\textrm{ch}}{\sqrt{1+\omega^2\tau^2_\textrm{ch}}} \tilde{\mathbf{E}}_\textrm{pr}(\omega)\cdot\mathbf{B}. \eea For $\mathbf{E}_\textrm{pr}\parallel\mathbf{B}$ the current density reads as

\bea J^{(1)}=-e \eta \int d\varepsilon~ v(\varepsilon)\nu(\varepsilon) f^{(1)}_{1}(\varepsilon)=\frac{\tau_\textrm{ch}}{1-i\omega\tau_\textrm{ch}} \frac{e^2v}{4\pi^2\hbar l_{B}^2} E, \eea yielding the following expression for the optical magneto-conductivity  
\bea \sigma_\textrm{ch}(\omega)=\frac{\tau_\textrm{ch}}{1-i\omega\tau_\textrm{ch}} \frac{e^2v}{4\pi^2\hbar l_{B}^2}\eea which, at the DC limit, reduces to the expression obtained in Ref. \cite{sonChiralAnomalyClassical2013c}. 
 
 

\subsection{Nonlinear optical response}


 The nonlinear response results from the $f_{1}^{(3)}$ in the distribution function. The linear dispersion of the chiral mode $\varepsilon_{0}(p_{z})=-\eta vp_{z}$ leads to constant density of states $\nu_{0}=1/2\pi l_{B}^2hv$, and the non-linear response vanishes identically. Therefore we have to depart from the linearly dispersed chiral mode and, specifically, we supplement it with quadratic dispersion. In Sec. \ref{lattice} we introduce a lattice model for Weyl semimetals where the lattice effects yield a quadratic correction to the linear dispersion with the following expression for the density of states: \bea \nu(\varepsilon)\approx \nu_0(1-\alpha\varepsilon/2+3\alpha^2\varepsilon^2/4), \eea 
where $\alpha^{-1}= mv^2/2$. For our cases $\alpha\varepsilon_{F}\!\ll\!1$ holds. We further assume that the scattering from the impurities relaxes the momentum. That is, the scattering rate is proportional to the density of states
$ \tau^{-1}(\varepsilon)=A \nu(\varepsilon)$, where the constant $A$ depends on the scattering strength.  
 
In the regime of interest $\omega\tau\gg 1$, the chiral number density $\delta n_{\eta}=\int d\varepsilon~ \nu(\varepsilon) f^{(3)}_{1}(\varepsilon)$ reads as 
 
 \bea \delta n_{\eta}=\eta \frac{15\alpha^2}{8\nu_{0}^2} \frac{1}{\omega^4\tau_\textrm{ch}} \frac{e^2\mathbf{E}_\textrm{pr}\cdot\mathbf{B}}{h^2}\left(\frac{e^2\mathbf{E}_\textrm{pu}\cdot\mathbf{B}}{h^2}\right)^2, \eea and for chiral density pumping we have 
 
 \bea \delta \tilde{n}(\omega)=\delta n_{+}-\delta n_{-}=\frac{15\alpha^2}{4\nu_{0}^2} \frac{1}{\omega^4\tau_\textrm{ch}} \frac{e^2\tilde{\mathbf{E}}_\textrm{pr}\cdot\mathbf{B}}{h^2}\left(\frac{e^2\tilde{\mathbf{E}}_\textrm{pu}\cdot\mathbf{B}}{h^2}\right)^2. \eea
 
We also obtain the following expression for the non-linear conductivity  

\bea \delta \sigma^{\mathrm{NL}}(\omega)=\frac{|e|^7v}{h^6}\frac{3\alpha^2\tau_\textrm{ch}^3}{2\nu^2_{0}} \frac{-1+7i\omega\tau_\textrm{ch}+10\omega^2\tau_\textrm{ch}^2-3i\omega^3\tau_\textrm{ch}^3}{(1-i\omega\tau_\textrm{ch})^4(1-2i\omega\tau_\textrm{ch})^2}B\left(\tilde{\mathbf{E}}_\textrm{pu}\cdot\mathbf{B}\right)^2, \eea
which clearly shows that the signal vanishes for perpendicular field alignments $\mathbf{E}_\textrm{pu}\perp\mathbf{B}$. Therefore the enhancement observed in the reflection results from the above nonlinear conductivity. In particular we see that the imaginary part is positive as 

\bea \mathrm{Im}[\delta\sigma^{\mathrm{NL}}(\omega)]=\frac{|e|^7v}{h^6}\frac{9\alpha^2}{8\nu^2_{0}} \frac{\tau_\textrm{ch}^3}{(\omega\tau_\textrm{ch})^3}B\left(\tilde{\mathbf{E}}_\textrm{pu}\cdot\mathbf{B}\right)^2,\eea  
which is required for the enhancement of reflection. This equation is the same as equation (4) in the main text.        


\section{A simple lattice model for Weyl semimetals \label{lattice}} 
We present a simple lattice model for a Weyl semimetal to simulate the essential features such as the isolated touching points between non-degenerate valence and conduction bands in the Brillouin zone. The lattice model reads as 

\bea H(\mathbf{k})=t[\cos (k_{x}a)+\cos (k_{y}a)-2]\sigma^{z}-t_{z}[\cos (k_{z}c)-\cos (Qc)]\sigma^{z}+t_{xy}\left[\sin(k_xa)\sigma^x+ \sin(k_ya)\sigma^y \right],\nonumber\\ \eea where $a$ and $c$ denote the lattice constants in the $ab$ plane and along the $c$ axis. The Weyl nodes are located at $k_z=\pm Q$: 

\bea \label{HW} H_{W}=\hbar v(k_{x}\sigma^{x}+k_{y}\sigma^{y}) - t_{z}[\cos (k_{z}c)-\cos (Qc)]\sigma^{z}, \eea where $v=t_{xy}a/\hbar$. In the presence of the quantizing field $\mathbf{B}=B\hat{z}$ the Landau levels are obtained via the substitution $\boldsymbol{\Pi}=\mathbf{p}+|e|\mathbf{A}$ for canonical momentum $\mathbf{p}=\hbar\mathbf{k}$, where in the Landau gauge $\mathbf{A}=(0,Bx,0)$. We use the following commutation relation for $\Pi_{\pm}=\Pi_{x}\pm i\Pi_{y}$
\bea [\Pi_{-},\Pi_{+}]=2e\hbar B  \eea to define lowering and raising boson operators 
\bea b=\frac{l_{B}}{\sqrt{2}\hbar}\Pi_{-},~~~~b^{\dag}=\frac{l_{B}}{\sqrt{2}\hbar}\Pi_{+}.  \eea The Weyl Hamiltonian (\ref{HW}) is written as 

\bea 
H_{W} = 
 \begin{pmatrix}
  \varepsilon_{0}(k_z) & \frac{\sqrt{2}\hbar v}{l_{B}}b \\
  \frac{\sqrt{2}\hbar v}{l_{B}}b^{\dag} & -\varepsilon_{0}(k_z) 
 \end{pmatrix},~~~~\varepsilon_{0}(k_z)=-t_{z}[\cos (k_{z}c)-\cos (Qc)], 
 \eea
 
 and the energy spectrum reads as
\[ \varepsilon_{n}(k_z)=
  \begin{cases}
    \pm\sqrt{\varepsilon_{0}^2(k_z) +\frac{2\hbar^2 v^2}{l^2_{B}}n }    & \quad n=1, 2, \cdots\\
     \varepsilon_{0}(k_z) & \quad n=0.
  \end{cases}
\] 

Near the node we have  
 
 \bea \varepsilon_{0}(k_z)\approx \frac{\hbar^2k_{z}^2}{2m}+\hbar v_{z}k_z,\eea
 where $m^{-1}=2t_{z}c^2\cos(Qc)/\hbar^2$, $v_{z}=t_{z}c\sin(Qc)/\hbar$, and $k_{z}$ is measured from the node $Q$.

\section{Discussion on the observed pump-probe signal  \label{PP_signal}}

\subsection{Sign of reflection pump-probe signal  \label{reflection}}

The permittivity $\epsilon$ can be written as follows in terms of the conductivity $\sigma(\omega)$,

\begin{equation} \label{eps}
\epsilon=\epsilon_\infty+i\frac{\sigma}{\omega\epsilon_0},
\end{equation}
and reflection is calculated as 

\begin{equation} \label{R}
R(T)=\left|\frac{1-\sqrt{\epsilon}}{1+\sqrt{\epsilon}}\right|^2
\end{equation}

For $|\Delta R| \ll R$ and $|\Delta \epsilon| \ll |\epsilon_b|$ ($\epsilon_b$: background permittivity), one can obtain the following relation for the change in reflection,

\begin{equation} \label{delR}
\Delta R \approx 2 |\epsilon_b|^{-3/2} |\Delta \epsilon|\cos(3\theta/2-\gamma) 
\end{equation}
where $\gamma= \angle \Delta \epsilon$ and  $\theta = \angle \epsilon_b$. 

From Fig. 1(a) in Ref.\cite{levyObservingOpticalSignaturesc}, we can estimate the permittivity of TaAs at $\omega=$14 meV to be $\epsilon_b=-530+i450=695 \angle 140^o$, and thus $\epsilon_b=695$ and $\theta=140^o$. From \eqref{delR}, in order to get a positive pump-probe signal ($\Delta R>0$), we need have a pump-induced change in reflection with $30^o<\gamma<210^o$. This means that nonlinear chiral charge pumping that gives rise to a large imaginary change in conductivity ($\gamma=90^o$) should lead to a positive change in reflection. Similarly, for hot carriers effects with a positive change in Drude weight ($\gamma=45^o$), $\Delta R$ will be positive. 

\subsection{Contribution of  Weyl bands to the  pump-probe response}
There are three sets of energy dispersion bands in TaAs: electron-doped W1 Weyl bands, electron-doped W2 Weyl bands, and  topologically-trivial hole-doped bands with a bandgap of $\approx$50 meV \cite{arnoldChiralWeylPockets2016b,levyObservingOpticalSignaturesc}. From optical reflection spectroscopy, the Fermi energy of W1, W2, and hole-doped bands are estimated to respectively be 14 meV, 2 meV, and 20 meV  for the samples considered in our study \cite{levyObservingOpticalSignaturesc}. Based on the numbers from TaAs bandstrucutre calculations \cite{lee2015fermi},  we estimate that  for $B \ge 2$ T, W2 Weyl bands  go to extreme quantum limit where only the zeroth Landau level is occupied. 

 In the following, we examine the contribution of different bands to the measured pump-probe response:
\subsubsection{Hot carriers response} 
The initial hot carriers response exhibits a positive change in reflection that is a signature of Drude weight increase. In the gapped hole band,  Sommerfeld expansion implies that  the chemical potential and hence the Drude weight of carriers  decreases with temperature. This behavior is similar to Drude weight dependence on temperature of semiconductors. Therefore, the observed Drude weight increase due to the hot carriers  has to be from W1 and W2 Weyl bands where hot carrier electron-hole pair production can increase the Drude weight. As we increase the magnetic field, W2 carriers go to extreme quantum limit where they no longer contribute to a temperature-dependent Drude weight. Therefore, for high magnetic fields, W1 carriers  only are responsible for the initial hot carriers response. This explains the behavior of $\Delta_1$ with magnetic field in Fig. 3c of the main part of the manuscript.

\subsubsection{Nonlinear chiral charge pumping response} 
In principle, both W1 and W2 Weyl nodes can contribute to a pump-induced chiral charge pumping processed discussed in our study. For W1 Weyl bands where many Landau levels are occupied, the pump-induced chiral quasiparticles (yellow carriers in Fig. 1c of the main text) can relax back to equilibrium via intra-Weyl-node relaxations. On the other hand, for W2 carriers residing in the extreme quantum limit, the pump-induced chiral quasiparticles can only be relaxed back to the equilibrium chemical potential by an chiral charge pumping (inter-Weyl-node) relaxation process. Therefore, only nonlinear chiral charge pumping in W2 bands can only contribute to the observed metastable state persisting  with a long time constant of $\tau_\textrm{ch}$.  

\subsection{Estimation of the pump-induced chiral charge pumping signal } 
From equations (2) and (4) in the main text, the ratio of nonlinear to linear chiral pumping conductivity can be written as 
\begin{equation} \label{ratio}
r_\textrm{NL}\equiv \left|\frac{\delta\sigma_{\textrm{ch}}^{\textrm{NL}}}{\sigma_{\textrm{ch}}}\right|=\frac{9}{8}\left(\frac{\alpha e v        }{\omega}\right)^2 |\tilde{E}_{\textrm{pump}}| ^2
\end{equation}
For W2 Weyl pocket, we have $v=2 \times 10^5$ m/s, and $\alpha=(50 \textrm{meV})^{-1}$ \cite{lee2015fermi,levyObservingOpticalSignaturesc}. Therefore, for  $\tilde{E}_{\textrm{pump}}$=50 kV/cm, and our measured frequency of $\hbar \omega=14$ meV, $r_\textrm{NL}\approx 1$. 

\subsection[scale=1.5]{Chiral charge pumping processes in the pump-probe measurements}
\begin{figure} [htbp]
 \centering
   \includegraphics[scale=1.5]{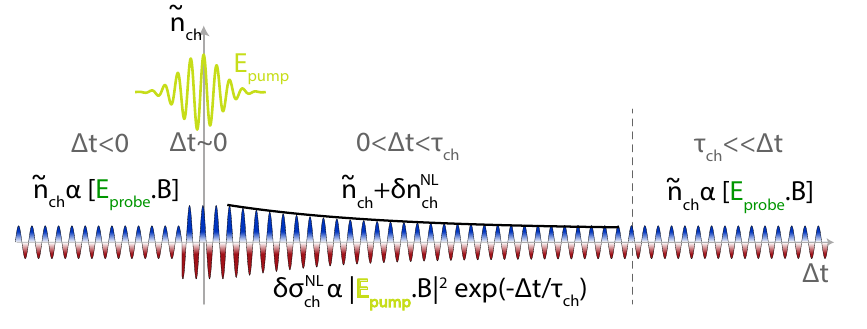}
  \caption{  (e) Schematic of the chiral charge density induced by the probe pulse at different  pump-probe time delays. For negative time delays ($\Delta t < 0$), the chiral current is linear with $E_\textrm{probe}$ and sinusoidally oscillates. For positive time delays, the strong pump pulse induce changes to the probe-produced chiral current. The nonlinear pump-induced change in the chiral current goes to zero at time delays much larger than the chiral pumping relaxation time $\Delta t \gg \tau_{ch}$. }
  \label{fig:S2}
\end{figure}

In order to provide insight into the dynamics of chiral carriers in the described  pump-probe measurements, in Fig.~\ref{fig:S2}, we sketch a cartoon image of the chiral charge density $\tilde{n}_{\textrm{ch}}$ produced by the probe pulse ($\mathbf{E}_{\textrm{probe}} \parallel \mathbf{B}$) as a function of the pump-probe time delay $\Delta t$. For $\Delta t <0$, i.e. before the pump pulse impinges on TaAs, the chiral charge imbalance  is $\tilde{n_{\textrm{ch}}}\propto\tilde{\sigma}_{\textrm{ch}} E_\textrm{probe}$   (from equation 2 in main text) and hence  oscillates synchronously with E$_\textrm{probe}$  (similar to Fig. 1b). At $\Delta t \sim 0$, when the pump pulse illuminates  the TaAs, the chiral anomaly conductivity seen by the probe pulse increases by $\delta \tilde{\sigma_{\textrm{ch}}}$ (from \eqref{eq:3}), causing $\tilde{n}_{\textrm{ch}}$ to enhance. The conductivity enhancement, which appears as an increase in the probe reflection, continues for $\Delta t > 0$ (i.e. after the pump pulse passed) as long as the quasiparticle excitations produced by the pump-induced chiral charge pumping persist. The picture is that the optical pump creates excited quasiparticles and that these excited quasiparticles modify the optical conductivity measured by the probe. As described before, this process is governed by the chiral pumping relaxation with time constant of $\tau_\textrm{ch}$. For  $\Delta t \gg \tau_\textrm{ch}$, $\tilde{n}_\textrm{ch}$ relaxes back to the equilibrium oscillations similar to the case for $\Delta t <0$. 

\section{Supplemental pump-probe data for $E_{\textrm{pump}}\perp B$}


\begin{figure} [htbp]
 \centering
   \includegraphics[scale=1.5]{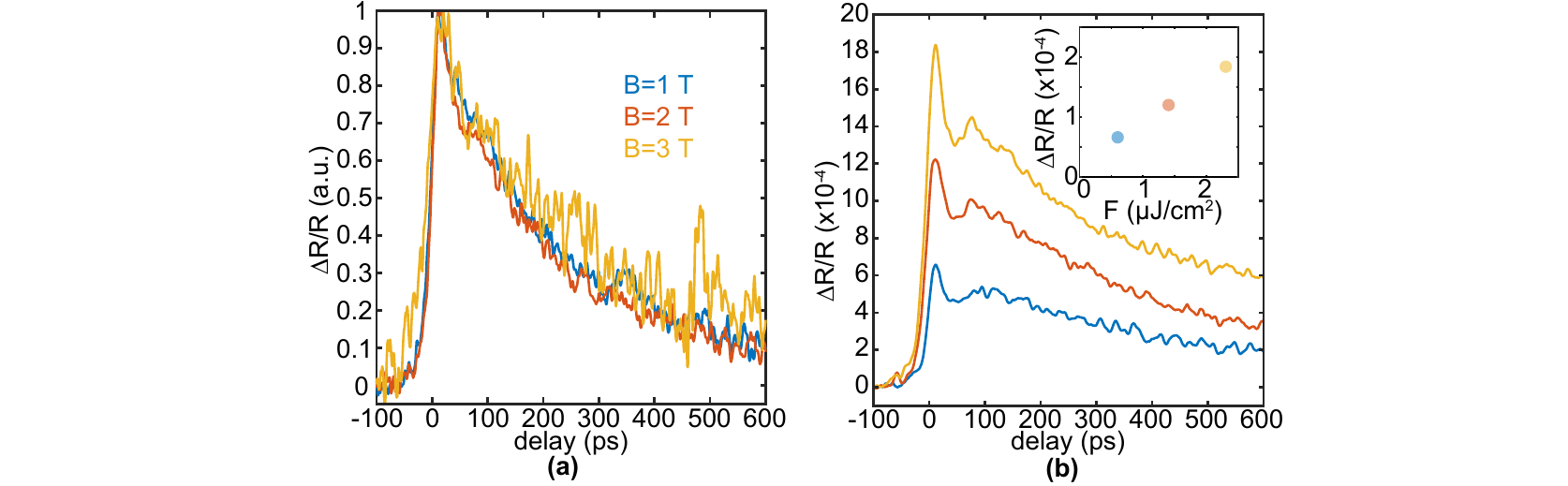}
  \caption{ (a) Pump-probe signal in arbitrary units at different magnetic fields. (b) Relative pump-induced increase in probe reflection for variety of pump fluences as a function of pump-probe time delay at $B=7$ T. All the pump-probe signals for $E_\textrm{pump}\perp E_\textrm{probe} \parallel B$ at different pump-fluences and magnetic fields exhibit similar time dynamics, and no long-lived response.} 
  \label{fig:S1}
\end{figure}

\end{document}